\documentclass[twocolumn]{aastex701}
\usepackage{amsmath}
\usepackage{nicefrac}
\usepackage[none]{hyphenat}

\def\Msun{M$_{\odot}$}

\begin{document}

\title{The Close Binary Properties of Massive Stars \\ across Different Environments within the LMC}

\author{Maxwell~Moe}
\affiliation{Department of Physics and Astronomy, University~of~Wyoming, 1000~E.~University~Ave., Dept.~3905, Laramie, WY 82071, USA}
\email{mmoe2@uwyo.edu}

\author{M.~S.~Oey}
\affiliation{Department of Astronomy, University of Michigan, 1085~S.~University~Ave., Ann~Arbor, MI, 48109, USA}
\email{msoey@umich.edu}

\author{Irene~Vargas-Salazar}
\affiliation{Department of Astronomy, University of Michigan, 1085~S.~University~Ave., Ann~Arbor, MI, 48109, USA}
\email{ivargasa@umich.edu}

\author{Kaitlin~M.~Kratter}
\affiliation{Steward Observatory, University of Arizona, 933~N.~Cherry~Ave.,~Tucson,~AZ 85721,~USA}
\email{kkratter@arizona.edu}

\begin{abstract}
We analyze 4,859 O-stars in the OGLE-III photometric survey of the LMC, including 415 eclipsing binaries (EBs). After accounting for the geometrical probability of eclipses, the period distribution of O-type binaries across $P$~=~2.5\,-\,200~days follows a power-law $f_{\rm logP}$~$\propto$~(log\,$P$)$^{\Pi}$ with $\Pi$~=~$-$0.34\,$\pm$\,0.06, which is skewed toward shorter periods compared to Opik's law ($\Pi$~=~0). We divide our O-stars into seven environments based on their clustering with B-stars and other O-stars. The EB fraction of O-stars in young clusters is 10.2\%\,$\pm$\,0.6\%, which matches the 10.8\%\,$\pm$\,2.1\% for O-stars in young Milky Way clusters. O-stars in old clusters exhibit a lower EB fraction of 5.5\%\,$\pm$\,0.9\% due to the effects of binary evolution. O-stars in young dense clusters, young sparse associations, and even low-mass clusters that formed in situ in the field have similar EB fractions. This uniformity suggests that the formation of close massive binaries depends on small-scale gas physics, e.g., fragmentation and migration within protostellar disks, whereas N-body interactions that scale with cluster density do not affect the close binary properties of massive stars that remain in clusters. Conversely, ejected O-stars in the field exhibit a lower close binary fraction. The EB fractions of field walkaways (projected velocities $v_{\rm proj}$~$<$~24.5~km~s$^{-1}$) and field runaways ($v_{\rm proj}$~$>$~24.5~km~s$^{-1}$) are 7.3\%\,$\pm$\,1.0\% and 4.7\%\,$\pm$\,1.0\%, respectively. These values suggest that most field O-stars were dynamically ejected via N-body interactions from their birth clusters, whereas field O-stars that formed in situ or were kicked from supernova explosions in binaries contribute 17\% and $<$\,28\%, respectively, to the field population. 
\end{abstract}


\section{Introduction}
\label{sec:Intro}

Most stars are born in binary or multiple systems (see \citealt{Offner2023} for a recent review). The close binary fraction (orbital separations $a$~$<$~10~au) strongly increases with primary mass, reaching 70\% for O-stars \citep{Sana2012,Kiminki2012}.  Massive protostellar disks are most prone to fragmentation, likely leading to the high occurrence rate of close companions to massive stars \citep{Kratter2006,Tokovinin2020}. 

The binary properties of O-stars vary with environment \citep{Gies1986,Gies1987,Mason2009,DorigoJones2020}. For example, \citet{Mason2009} combined previous spectroscopic surveys with their own high-angular resolution observations to perform a relatively complete multiplicity census of Galactic O-stars in three different environments. They measured a binary fraction of 75\% for O-stars in clusters and associations, but only 59\% for field O-stars far removed from any clusters. For runaway O-stars, traditionally defined as those having peculiar velocities exceeding $v$~$>$~30~km~s$^{-1}$ \citep{Gies1987,Hoogerwerf2001}, \citet{Mason2009} measured an even lower binary fraction of 43\% (see their Table~5). 

There are two preferred mechanisms for producing runaway O-stars: N-body interactions that dynamically eject O-stars from their birth clusters and kicks in massive binaries when one of the components undergoes a supernova explosion \citep{Blaauw1961,Gies1986,Gies1987,Stone1991,Hoogerwerf2001,deWit2005,Perets2012,Oh2016}. However, recent modeling suggests that most supernova explosions disrupt massive binaries with slower speeds $v$~$<$~30~km~s$^{-1}$ \citep{Renzo2019}.  Ejected O-stars that do not meet the runaway threshold of $v$~$>$~30~km~s$^{-1}$ are called walkaways. 

Variations in the close binary fraction of O-stars across different environments provide invaluable insight into the formation process and dynamical evolution of massive stars. The goal of this study is to measure the eclipsing binary (EB) fraction of O-stars across different regions within the LMC. The third phase of the Optical Gravitational Lensing Experiment (OGLE-III) photometrically monitored 35 million stars in the LMC and identified 26,121 EBs \citep{Udalski2008,Graczyk2011}. The OGLE-III LMC sample of massive stars is more than an order of magnitude larger than Galactic surveys, allowing us to divide LMC O-stars into several different environments while retaining informative EB statistics. In Section~\ref{sec:sample}, we discuss our selection criteria and overall properties of our sample. We compare the EB statistics of O-stars in clusters and in the field in Sections~\ref{sec:cluster} and \ref{sec:field}, respectively. We summarize our key results in Section~\ref{sec:summary}.

\section{Sample Selection}
\label{sec:sample}
\subsection{Photometric O-Star Sample}

\begin{figure}[t!]
\centerline{
\includegraphics[trim=1.4cm 0.1cm 1.5cm 0.1cm, clip=true, width=3.4in]{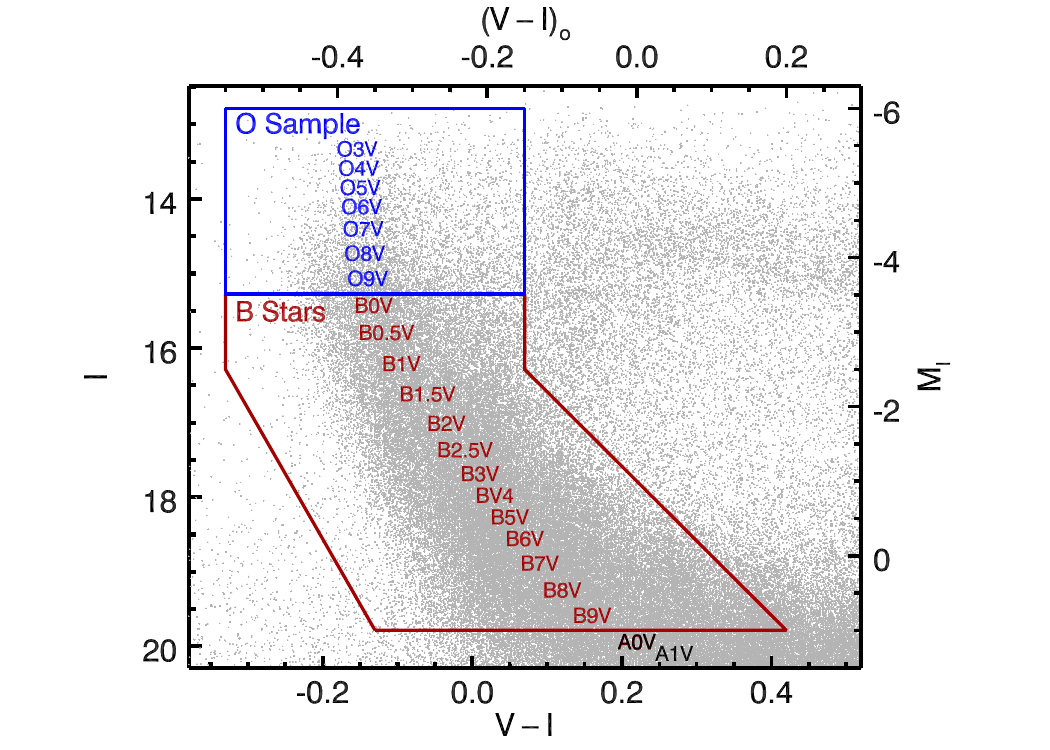}}
\caption{Color-magnitude diagram of blue, luminous stars in the OGLE-III LMC survey. Our photometric sample contains the 4,859 photometric O-stars (blue rectangle). We use the 2.23 million photometric B-dwarfs (red region) to differentiate the ages and environments of our O-stars. }
\label{fig:HRD}
\end{figure}

For seven years, OGLE-III monitored 35 million stars in the LMC in both V and I bands \citep{Udalski2008}. We downloaded their photometric data from the 115 OGLE-III LMC fields\footnote{http://ogle.astrouw.edu.pl/}. Some stars have repeated entries in slightly overlapping fields. We therefore compare the positions and magnitudes of all bright, blue stars (I~$\le$~20.5~mag, V$-$I $\le$ 0.6~mag), and we eliminate repeated entries within 1'' that have magnitude differences less than $\Delta$I $\le$ 0.5~mag (6.9\% of the sample).  We adopt a distance modulus of $\mu$ = 18.48 ($d$ = 49.6~kpc) to the LMC \citep{Pietrzynski2019}, yielding a projected scale of 0.24~pc/''. We also adopt an average dust extinction of $\langle$A$_{\rm I}\rangle$ = 0.30~mag and dust reddening of $\langle$E(V$-$I)$\rangle$ = 0.22~mag appropriate for hot OB stars in the LMC \citep{Zaritsky2004,Moe2015a}. In Fig.~\ref{fig:HRD}, we display absolute magnitudes M$_{\rm I}$ = I\,$-$\,$\mu$\,$-$\,$\langle$A$_{\rm I}\rangle$ versus intrinsic colors (V$-$I)$_{\rm o}$ = (V$-$I)\,$-$\,$\langle$E(V$-$I)$\rangle$ for our bright, blue LMC stars. We indicate the spectral types of dwarf stars from the empirical relations presented in \citet{Pecaut2013}\footnote{updated at http://www.pas.rochester.edu/$\sim$emamajek/spt/}. O-stars are in the Rayleigh-Jeans tail with intrinsic color (V$-$I)$_{\rm o}$ = $-$0.37 while B-dwarfs become progressively redder toward later subtypes and fainter magnitudes. 

We initially select the 4,835 blue, luminous stars bounded by $-$6.0~$<$~M$_{\rm I}$~$<$ $-$3.5 and $-$0.55~$<$~(V$-$I)$_{\rm o}$~$<$~$-$0.15 as candidate O-stars (blue rectangle in Fig.~\ref{fig:HRD}). We add 24 EBs with mean OGLE-III magnitudes $-$3.5 $<$ $\langle$M$_{\rm I}\rangle$ $<$ $-$3.4 that are slightly fainter than our selection criterion but with out-of-eclipse magnitudes as reported in \citet{Graczyk2011} that are actually brighter than our limit. Our final sample includes 4,859 candidate O-stars determined from photometry. There are 2.23 million photometric B-stars (red region in Fig.~\ref{fig:HRD}), yielding a ratio of N$_{\rm B}$/N$_{\rm O}$ = 460. 

We acknowledge that such a photometric sample is neither complete nor pure. Late O-stars with larger than average dust extinction will appear fainter than our magnitude limit, and those in extremely crowded regions are unresolved in the ground-based OGLE-III photometry. Conversely, some early B-dwarfs with less than average dust extinction will likely leak into our photometric O-star sample. Nonetheless, the precise boundary separating O versus B-stars is not important for our purposes. By applying the same photometric criteria across the entire LMC, the homogeneity of the OGLE-III survey allows us to measure the EB fractions of massive stars across a wide range of different environments.

\subsection{Overall Eclipsing Binary Fraction}

Of our 4,859 O-stars, OGLE-III obtained N$_{\rm I}$ $\ge$\,120 I-band measurements for 4,814 (99.1\%) of them and subsequently searched their light curves for EB companions \citep{Graczyk2011}. We consider all 4,859 O-stars when determining their environments, but include only the 4,814 O-stars with N$_{\rm I}$ $\ge$\,120 I-band measurements when computing EB statistics. Within our O-star sample, \citet{Graczyk2011} identified 415 EBs, yielding an \textsc{Overall} EB fraction of F$_{\rm EB}$~=~8.6\%\,$\pm$\,0.4\%. We report this statistic at the top of the flowchart in Fig.~\ref{fig:flowchart} and in Table~\ref{tab:Ostars}.

\citet{Sana2012} thoroughly investigated the close binary properties of 71 O-stars in young Galactic clusters, which provides a benchmark for comparison. They identified 40 spectroscopic binaries, including 23 with short periods $P$ $<$ 20 days. In their sample, 9 are EBs, yielding an observed EB fraction of F$_{\rm EB}$ = 12.7\%\,$\pm$\,3.9\%. To reduce the uncertainty, we utilize the full population of spectroscopic binaries under the assumption that they can be viewed with random orientations. We adopt the spectral types and orbital properties of their spectroscopic binaries (see Table~S4 in \citealt{Sana2012}). We map the spectral types to stellar masses $M$ and radii $R$ according to the empirical relations in \citet{Pecaut2013}. EBs are skewed toward very short orbital periods and thus are tidally dissipated to small eccentricities. For circular orbits, the probability of observing an eclipse is:

\begin{equation}
 p_{\rm EB} = \frac{R_1 + f\,R_2}{a}, 
 \label{eqn:prob}
\end{equation}

\noindent  where $f$ = $-$1 corresponds to a non-grazing, full transit and $f$ = 1 corresponds to a barely grazing eclipse. Given the precision of the photometric monitoring of nearby Galactic O-stars, we adopt $f$ = 0.8. By adding the eclipse probabilities for all spectroscopic binaries, we compute an expected EB fraction of F$_{\rm EB}$ = 10.8\%\,$\pm$\,2.1\%. The uncertainty derives from adding in quadrature the binomial error and a 10\% uncertainty in the radii. The expected EB fraction is consistent with the measured value, and we list both benchmarks at the top of Fig.~\ref{fig:flowchart}. The \textsc{Overall} LMC O-star EB fraction of 8.6\%\,$\pm$\,0.4\% is slightly smaller than but consistent with the value for O-stars in young Milky Way clusters (see also Section~\ref{sec:cluster}).

\begin{figure}[t!]
\centerline{
\includegraphics[trim=1.5cm 0.1cm 0.3cm 0.1cm, clip=true, width=3.4in]{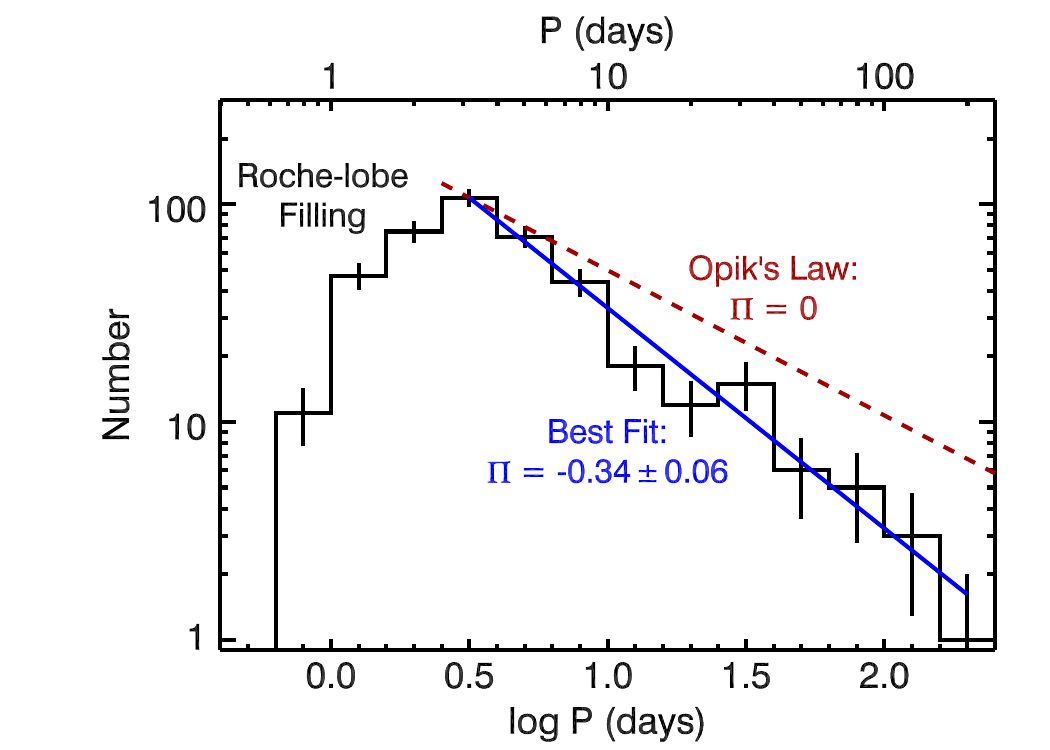}}
\caption{Orbital period distribution of LMC O-type EBs. Only compact, young O-dwarfs can host EB companions below $P$~$<$~2~days without overfilling their Roche lobes. After accounting for the geometrical probability of eclipses, the underling period distribution of O-type binaries across $P$~=~2.5\,-\,200~days can be accurately modeled by a power-law $f_{\rm logP}$~$\propto$~(log,$P$)$^{\Pi}$ with $\Pi$~=~$-$0.34\,$\pm$\,0.06 (blue), which is skewed toward shorter periods compared to Opik's law ($\Pi$~=~0) of a uniform distribution with respect to log $P$ (dashed red).  }
\label{fig:Pdist}
\end{figure}

We display in Fig.~\ref{fig:Pdist} the period distribution of our 415 LMC O-type EBs. The distribution peaks near $P$~=~3~days with an extended tail toward longer periods. We report in Table~\ref{tab:Ostars} the median orbital period of $P$~=~3.49\,$\pm$\,0.11~days, where the uncertainty derives from simulating 1,000 bootstrap samples. Binaries below $P$~$<$~2~days must contain young, compact O-dwarf primaries that avoid overfilling their Roche lobes. Across $P$ = 2.5\,-\,200 days, we fit a power-law distribution $f_{\rm logP,EB}$ $\propto$ (log\,$P$)$^\gamma$ with $\gamma$~=~$-$1.01\,$\pm$\,0.06. Given that the geometrical probability of eclipses scales as $p_{\rm EB}$ $\propto$ $a^{-1}$ $\propto$ $P^{-2/3}$ (Eqn.~\ref{eqn:prob}), then the intrinsic period distribution of O-type binaries is $f_{\rm logP}$~$\propto$~(log\,$P$)$^\Pi$ with $\Pi$ = $\gamma$ + 0.67 = $-$0.34\,$\pm$\,0.06. O-type binaries are skewed toward shorter periods compared to Opik's law with $\Pi$~=~0 (uniform in log $P$) or even early-B binaries with $\Pi$~=~$-$0.2 \citep{Moe2013}. The measured distribution is slightly shallower than but consistent with the power-law slope of $\Pi$~=~$-$0.55\,$\pm$\,0.22 measured from O-type spectroscopic binaries across a slightly broader range of orbital periods log\,$P$\,(days) = 0.15\,-\,3.5 \citep{Sana2012}.

\section{Cluster O-stars}
\label{sec:cluster}

\begin{figure}[t!]
\centerline{
\includegraphics[trim=0.4cm 0.1cm 0.3cm 0.1cm, clip=true, width=3.4in]{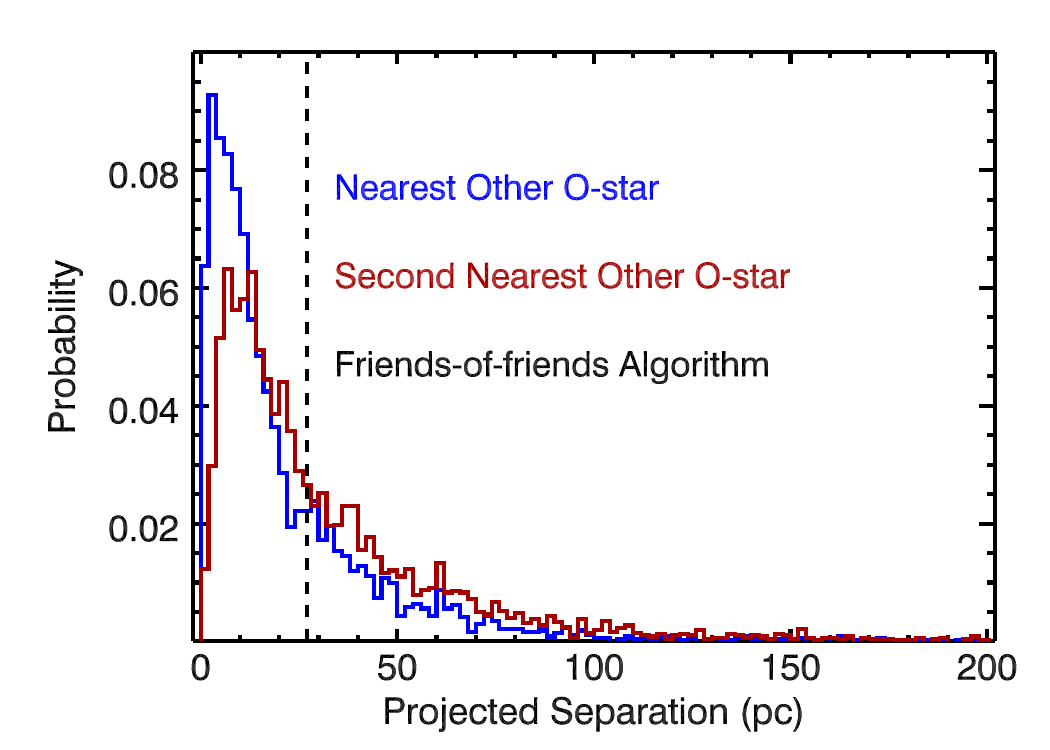}}
\caption{Probability distributions of projected separations between LMC O-stars and nearest other O-star (blue) and second nearest other O-star (red). A friends-of-friends algorithm determines a representative clustering length between O-stars of 27 pc (dashed black). }
\label{fig:Ostarsep}
\end{figure}

We separate our LMC O-stars according to their environments, specifically their clustering with respect to other O-stars. In Fig.~\ref{fig:Ostarsep}, we display the distributions of projected separations between O-stars and the nearest and second nearest other O-stars. The average separations to the nearest and second nearest other O-stars are 22 pc and 32 pc, respectively. Using a friends-of-friends algorithm \citep{Battinelli1991,Vargas-Salazar2020}, we measure a clustering length between O-stars of 27 pc = 112'', halfway between the two computed averages above. Our clustering length of 27 pc for O-stars in the LMC is similar to the 28 pc clustering length for OB stars in the SMC \citep{Oey2004,Lamb2016}.

\begin{figure*}[t!]
\centerline{
\includegraphics[trim=0.4cm 1.6cm 1.5cm 0.2cm, clip=true, width=7.2in]{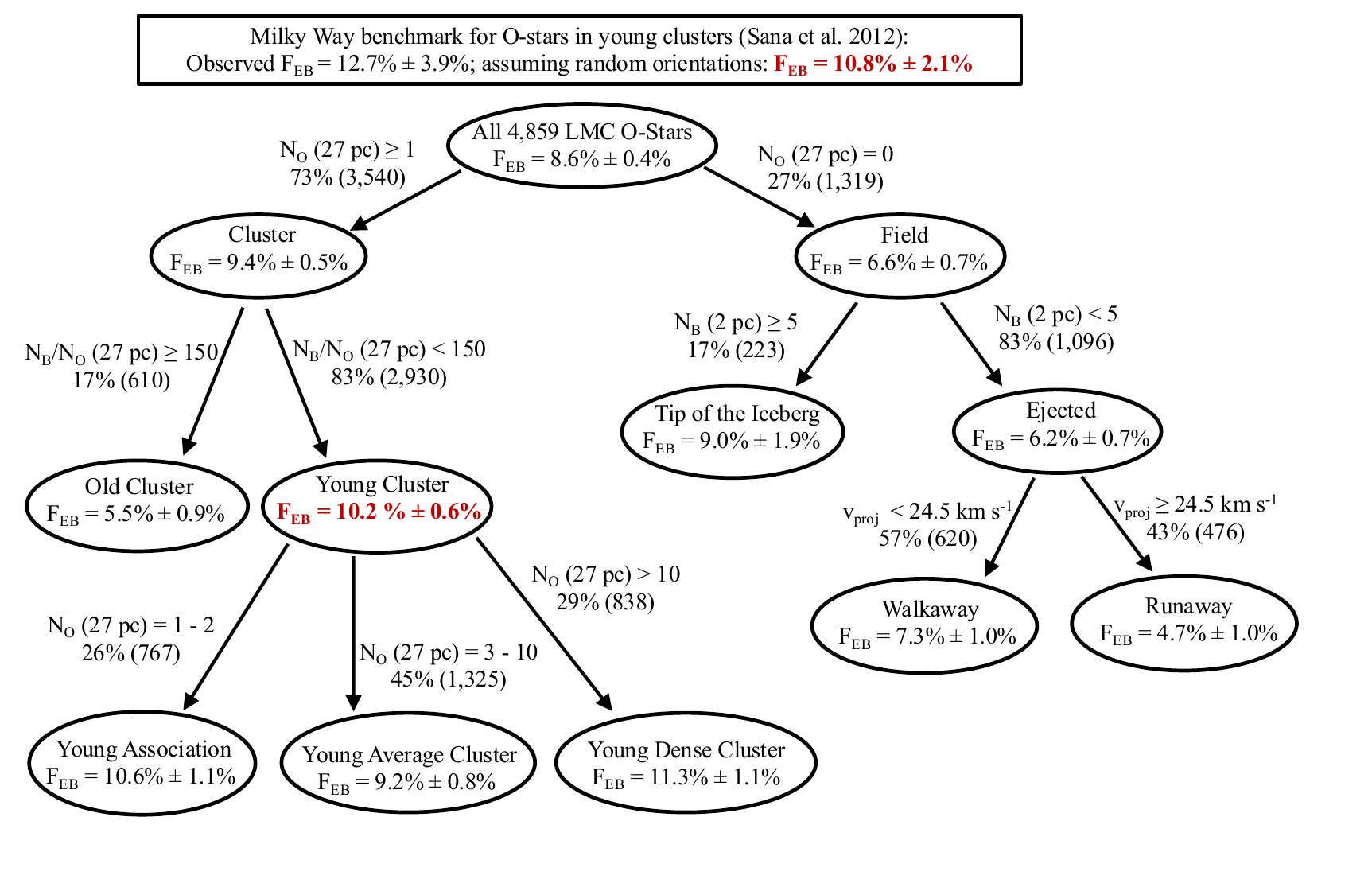}}
\caption{Classification of OGLE-III LMC O-stars according to their environments, their respective branching ratios, and their EB fractions. The EB fraction of LMC O-stars in \textsc{Young Clusters} matches the benchmark value of O-stars in young Milky Way clusters (bold red). The EB fraction is consistent for O-stars born in \textsc{Young Dense Clusters}, \textsc{Young Average Clusters}, sparse \textsc{Young Associations}, and even low-mass \textsc{Tip of the Iceberg} clusters that formed in situ within the \textsc{Field}. Meanwhile, O-stars in \textsc{Old Clusters} exhibit a reduced EB fraction due to the effects of binary evolution. Similarly, \textsc{Ejected} O-stars in the \textsc{Field} exhibit a lower EB fraction, including \textsc{Walkaway} and especially \textsc{Runaway} O-stars.}
\label{fig:flowchart}
\end{figure*}

We define the 3,540 O-stars (73\%) with at least one other O-star within a projected separation of $r_{\rm proj}$~$\le$~27~pc as \textsc{Cluster} O-stars. Meanwhile, we classify the remaining 1,319 O-stars (27\%) as \textsc{Field} O-stars. We display this bifurcation toward the top of the flowchart in Fig.~\ref{fig:flowchart} and present their properties in Table~\ref{tab:Ostars}.  Our \textsc{Cluster} O-stars include O-stars in canonical clusters and in sparse associations (see below). The EB fraction of \textsc{Cluster} O-stars is $F_{\rm EB}$ = 9.4\%\,$\pm$\,0.5\%, slightly larger than but consistent with the \textsc{Overall} value of 8.6\%\,$\pm$\,0.4\%. We discuss our \textsc{Field} O-star subset in Section~\ref{sec:field}.

\begin{deluxetable*}{ccccccc}[th!]
    \tablecaption{Properties of OGLE-III Photometric O-stars in the LMC}    \label{tab:Ostars}
    \startdata
      & & & & & & \\
     \rule{0pt}{3ex}Sample & N & F$_{\rm O}$ & N$_{\rm I}$ $>$ 120 & N$_{\rm EB}$ & F$_{\rm EB}$ & Median $P$ (days)\\
     \hline
     \rule{0pt}{3ex}\textsc{Overall}               & 4,859 & 100\% & 4,814 & 415 &  8.6\%\,$\pm$\,0.4\% & 3.49\,$\pm$\,0.11 \\
     \rule{0pt}{3ex}\textsc{Cluster}               & 3,540 &  73\% & 3,503 & 328 &  9.4\%\,$\pm$\,0.5\% & 3.43\,$\pm$\,0.13 \\
     \rule{0pt}{3ex}\textsc{Field}                 & 1,319 &  27\% & 1,311 &  87 &  6.6\%\,$\pm$\,0.7\% & 3.61\,$\pm$\,0.30 \\
     \rule{0pt}{3ex}\textsc{Old Cluster}             & 610 &  13\% &   601 &  33 &  5.5\%\,$\pm$\,0.9\% & 4.18\,$\pm$\,0.28 \\
     \rule{0pt}{3ex}\textsc{Young Cluster}         & 2,930 &  60\% & 2,902 & 295 & 10.2\%\,$\pm$\,0.6\% & 3.32\,$\pm$\,0.13 \\
     \rule{0pt}{3ex}\textsc{Young Dense Cluster}     & 838 &  17\% &   831 &  94 & 11.3\%\,$\pm$\,1.1\% & 3.43\,$\pm$\,0.27 \\
     \rule{0pt}{3ex}\textsc{Young Average Cluster} & 1,325 &  27\% & 1,313 & 121 &  9.2\%\,$\pm$\,0.8\% & 3.52\,$\pm$\,0.22 \\
     \rule{0pt}{3ex}\textsc{Young Association}       & 767 &  16\% &   758 &  80 & 10.6\%\,$\pm$\,1.1\% & 2.94\,$\pm$\,0.25 \\
     \rule{0pt}{3ex}\textsc{Tip of the Iceberg}      & 223 &   5\% &   223 &  20 &  9.0\%\,$\pm$\,1.9\% & 4.92\,$\pm$\,1.11 \\
     \rule{0pt}{3ex}\textsc{Ejected}               & 1,096 &  23\% & 1,088 &  67 &  6.2\%\,$\pm$\,0.7\% & 3.49\,$\pm$\,0.24 \\
     \rule{0pt}{3ex}\textsc{Walkaway}              &   620 &  13\% &   616 &  45 &  7.3\%\,$\pm$\,1.0\% & 3.74\,$\pm$\,0.27 \\
     \rule{0pt}{3ex}\textsc{Runaway}               &   476 &  10\% &   472 &  22 &  4.7\%\,$\pm$\,1.0\% & 3.26\,$\pm$\,0.28 \\
    \enddata
\end{deluxetable*}

\begin{figure*}[t!]
\centerline{
\includegraphics[trim=0.7cm 0.8cm 0.7cm 1.0cm, clip=true, width=6.7in]{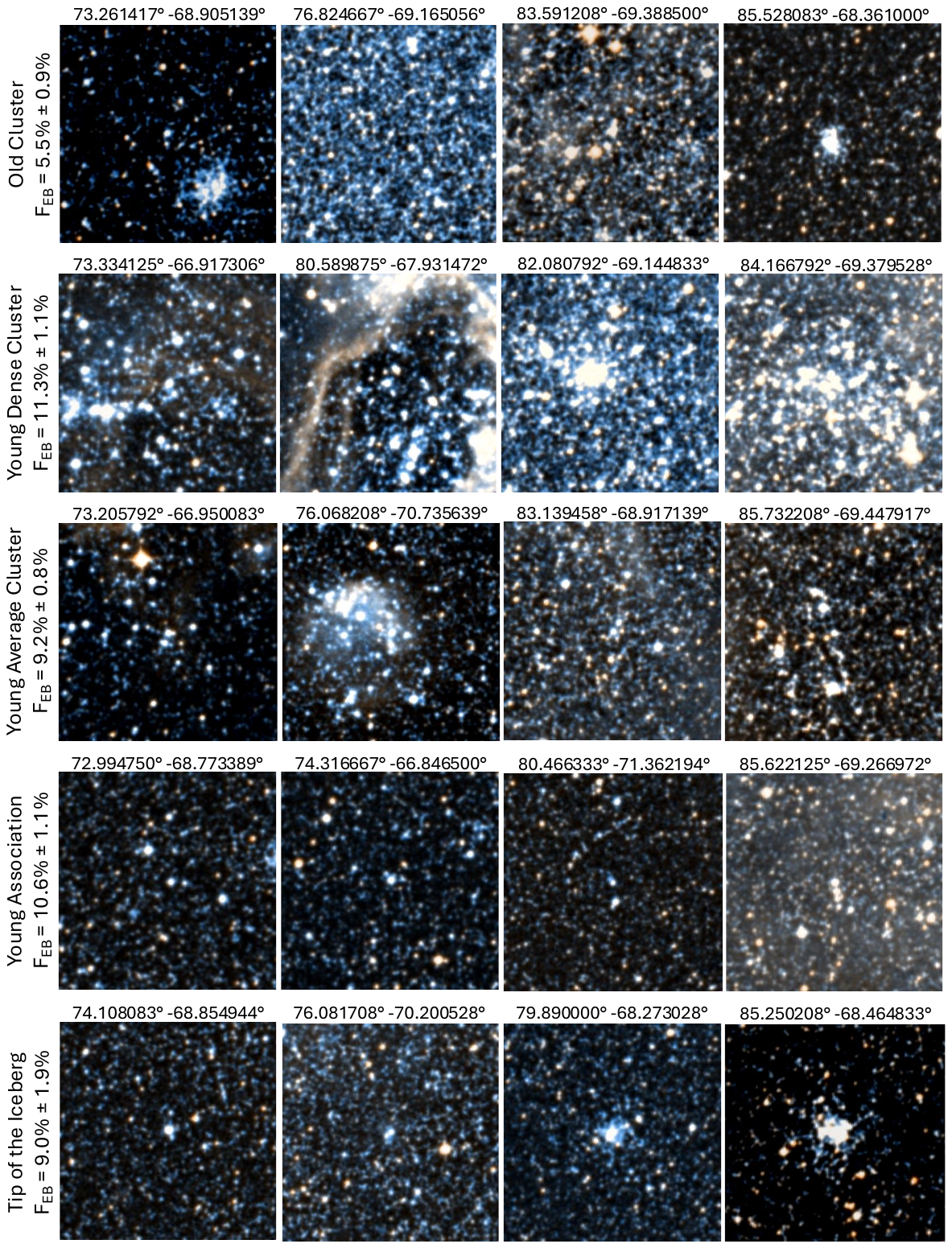}}
\caption{Example 50-pc\,$\times$\,50-pc DSS images and listed coordinates for selected O-stars according to their environments. O-stars in \textsc{Old Clusters} (top row) exhibit a lower EB fraction due to the effects of binary evolution. Meanwhile, the EB fractions of O-stars in different types of  \textsc{Young Clusters} (middle three rows) do not vary with respect to the stellar density of the cluster. Even O-stars that formed in isolated, low-mass \textsc{Tip of the Iceberg} clusters (bottom row) exhibit an EB fraction consistent with those born in \textsc{Young Clusters}. }
\label{fig:clusters}
\end{figure*}

We next utilize the ratio N$_{\rm B}$/N$_{\rm O}$ between the numbers of B and O-stars within $r_{\rm proj}$ $<$ 27~pc as an indicator of cluster age. Older populations will have relatively fewer O-stars since they have evolved into red supergiants and compact remnants.  The 610 (17\%) \textsc{Cluster} O-stars with $N_{\rm B}$/$N_{\rm O}$ $\ge$ 150 within $r_{\rm proj}$ $<$ 27~pc reside in \textsc{Old Clusters}, while the remaining 2,930 (83\%) \textsc{Cluster} O-stars are in \textsc{Young Clusters} (see Fig.~\ref{fig:flowchart} and Table~\ref{tab:Ostars}). In the top row of Fig.~\ref{fig:flowchart}, we show 50-pc\,$\times$\,\,50-pc DSS images from Aladin Lite\footnote{https://aladin.cds.unistra.fr/AladinLite/} centered on example O-stars in \textsc{Old Clusters}. A sizable fraction of old O-stars reside in or near old clusters as demonstrated by the enhanced clustering of nearby B-dwarfs, B-supergiants, and red supergiants. Some old O-stars are located in the central bar of the LMC (e.g., second example in top row of Fig.~\ref{fig:clusters}). The LMC bar is 60\,-\,250 Myr old, i.e., older than the lifetime of single O-stars, and hosts a significant population of B-dwarfs \citep{Nayak2016}. O-stars that linger in the LMC bar are possibly massive blue stragglers, i.e., the rejuvenated mass gainers in post-mass-transfer binaries \citep{Schneider2014}. Other O-stars in \textsc{Old Clusters} may have been ejected from nearby clusters into the LMC bar but still remain within 27~pc of at least one other O-star. In any case, they are systematically older than \textsc{Young Cluster} O-stars embedded in regions with $N_{\rm B}$/$N_{\rm O}$ $<$ 150.

The EB fractions of O-stars in \textsc{Old Clusters} versus \textsc{Young Clusters} are 5.5\%\,$\pm$\,0.9\% and 10.2\%\,$\pm$\,0.6\%, respectively, which are discrepant with each other at the 4.3$\sigma$ level. A larger fraction of close binaries in older environments have either merged or their primaries have evolved into compact objects, resulting in a smaller EB fraction. Given a constant star formation rate, \citet{deMink2014} estimated that 30\% of massive stars are the products of binary evolution. Similarly, \citet{Moe2017} computed that 20\% of OB stars are the secondaries in which the original primaries have already evolved into compact remnants. The EB fraction of O-stars in \textsc{Young Clusters} is 19\% higher than the \textsc{Overall} value, consistent with the \citet{Moe2017} estimate.

The median orbital periods of the \textsc{Old Cluster} and \textsc{Young Cluster} EBs are 4.18\,$\pm$\,0.28 days and 3.32 \,$\pm$\,0.13 days, respectively, which are discrepant with each other at the 3.2$\sigma$ level. Older O-stars are physically larger and cannot host very close companions without overfilling their Roche lobes. The peak in the period distribution of older EBs is therefore shifted toward longer periods. Nonetheless, the power-law slope of the long-period tail does not noticeably change. We measure $\Pi$ = $-$0.22\,$\pm$\,0.18 and $-$0.34\,$\pm$\,0.07 for the \textsc{Old Cluster} and \textsc{Young Cluster} EBs, respectively.

The EB fraction of 10.2\%\,$\pm$\,0.6\% for LMC O-stars in \textsc{Young Clusters} matches the benchmark value of 10.8\%\,$\pm$\,2.1\% for Milky Way young clusters \citep[][see Section~\ref{sec:sample}]{Sana2012}. The close binary fraction of massive stars does not vary between the LMC and Milky Way, consistent with  previous studies \citep{Moe2013,Sana2013,Dunstall2015,Almeida2017,Villasenor2021}. The slight metallicity difference between Galactic O-stars ([Fe/H] = 0.0) and LMC O-stars ([Fe/H] = $-$0.4) does not measurably affect their close binary properties.

We next separate the O-stars in young clusters according to cluster density, specifically the number of other O-stars within $r_{\rm proj}$ $<$ 27 pc. We consider the 767 (26\%) young O-stars with only one or two other O-stars within 27~pc to be members of \textsc{Young Associations}. Meanwhile, the 1,325 (45\%) with 3\,-\,10 other O-stars within 27~pc reside in \textsc{Young Average Clusters}, and the remaining 838 (29\%) with $>$\,10 other O-stars within 27~pc are members of \textsc{Young Dense Clusters} (see Fig.~\ref{fig:flowchart} and Table~\ref{tab:Ostars}). In the middle three rows of Fig.~\ref{fig:clusters}, we show DSS images of example O-stars in \textsc{Young Dense Clusters}, \textsc{Young Average Clusters}, and \textsc{Young Associations} from top to bottom. The \textsc{Young Dense Clusters} are typically massive clusters that extend beyond the 50-pc\,$\times$\,50-pc field of view, some with visible nebulosity. The \textsc{Young Average Clusters} tend to be more compact with radii $<$\,20~pc. The \textsc{Young Associations} are rather sparse and typically exhibit only minor clustering with one or two other O-stars in the field of view.

The EB fractions of O-stars in \textsc{Young Dense Clusters}, \textsc{Young Average Clusters}, and \textsc{Young Associations} are 11.3\%\,$\pm$\,1.1\%, 9.2\%\,$\pm$\,0.8\%, and 10.6\,$\pm$\,1.1\%, respectively, which are all consistent with each other (see Fig.~\ref{fig:flowchart}). Their median orbital periods exhibit slightly more scatter but are still consistent with each other at the 1.7$\sigma$ level (see Table~\ref{tab:Ostars}). Whether O-stars reside in dense clusters or sparse associations, their close binary fraction is independent of the surrounding stellar density.

 Close binary stars within $a$~$<$~10~au cannot form in situ \citep{Boss1986, Bate1998, Bate2002}. The components instead fragment at wider separations and migrate inward via N-body dynamical interactions in their birth cluster \citep{Heggie1975,Kroupa1995,Bate1998,Bate2009}, secular evolution with a tertiary \citep{Kiseleva1998,Naoz2014}, dissipation of orbital energy into the surrounding gaseous envelope or protostellar disk \citep{Artymowicz1991,Bate1995,Bate1997,Tokovinin2020}, or some combination thereof \citep{Bate2002,Moe2018,RamirezTannus2021}. The uniformity of the EB fraction of O-stars across a broad range of cluster densities suggests that the formation of close companions to massive stars depends solely on small-scale gas physics, e.g., fragmentation and migration within protostellar disks \citep{Tokovinin2020}. Meanwhile, N-body interactions on larger scales play a negligible role in shaping the close binary properties of \textsc{Cluster} O-stars.

\section{Field O-stars}
\label{sec:field}

We return to the 1,319 \textsc{Field} O-stars defined by not having any other O-stars within $r_{\rm proj}$ $<$ 27~pc. Their EB fraction is F$_{\rm EB}$ = 6.6\%\,$\pm$\,0.7\%, which is less than the \textsc{Cluster} value of 9.4\%\,$\pm$\,0.5\%, statistically significant at the 3.3$\sigma$ level (see Fig.~\ref{fig:flowchart} and below). Field O-stars consist of both O-stars that were \textsc{Ejected} far from their birth clusters, either through N-body interactions or supernova kicks in binaries, and O-stars that formed in situ within small, isolated clusters. We refer to a field O-star that formed in situ as \textsc{Tip of the Iceberg} considering that it represents the tip, i.e., the only O-star, in an underlying low-mass cluster \citep{Lamb2016,Vargas-Salazar2020}. 

\begin{figure}[t!]
\centerline{
\includegraphics[trim=0.4cm 0.1cm 0.3cm 0.1cm, clip=true, width=3.4in]{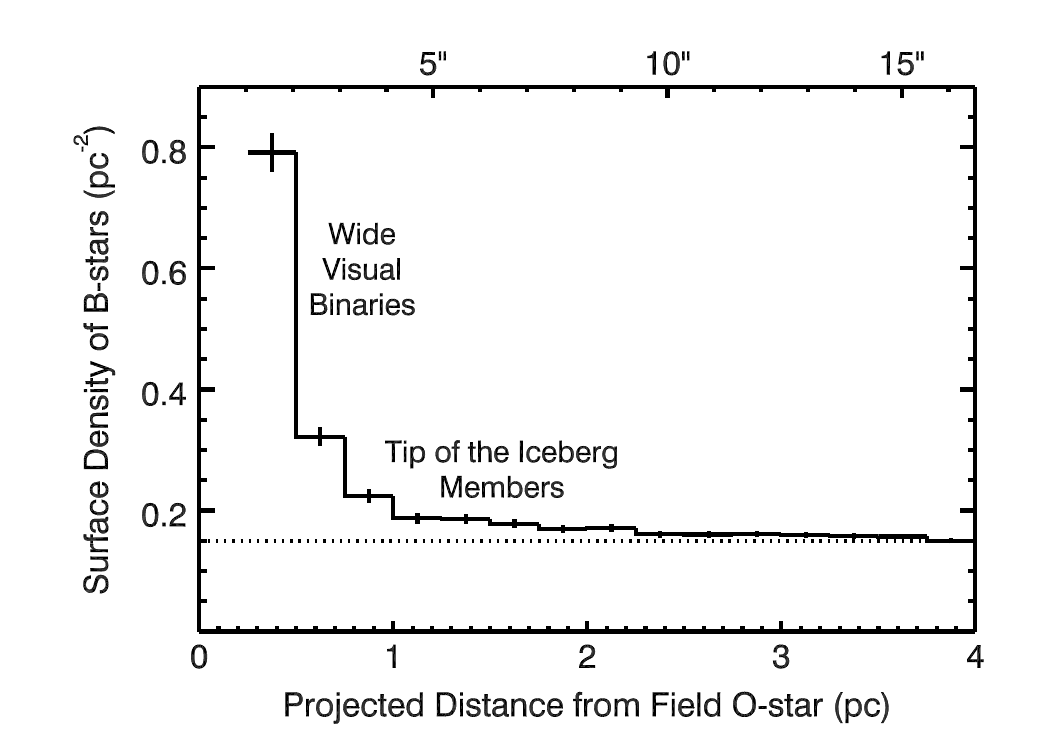}}
\caption{Average surface density of B-stars as a function of projected separation from our 1,319 \textsc{Field} O-stars. Spatially resolved visual binaries yield an enhancement within $r_{\rm proj}$~$<$~1~pc ($<$\,4''). \textsc{Tip of the Iceberg} O-stars in isolated low-mass clusters exhibit a measurable overdensity of B-stars across $r_{\rm proj}$ = 1\,-\,2 pc (4''\,-\,8''). }
\label{fig:Bdensity}
\end{figure}

We distinguish \textsc{Ejected} O-stars from \textsc{Tip of the Iceberg} O-stars according to the immediately surrounding density of B-stars. In Fig.~\ref{fig:Bdensity}, we display the surface density of B-stars as a function of projected separation from \textsc{Field} O-stars averaged across all 1,319 systems. Within $r_{\rm proj}$~$<$~1~pc ($<$\,4''), there is a substantial enhancement of B-stars that are wide, spatially resolved visual companions. Across $r_{\rm proj}$~=~1\,-\,2~pc (4''\,-\,8''), we measure a surface density of 0.18~B-stars~pc$^{-2}$, which is noticeably higher than the average background level of 0.15~B-stars~pc$^{-2}$. Such wide pairs across 1\,-\,2~pc cannot be orbiting visual binaries but instead signify common membership in low-mass \textsc{Tip of the Iceberg} clusters.

For a zero-age main-sequence (MS) population, we expect 13 B-stars with $M$~=~2.5\,-\,17\,\Msun\ for every O-star above $M$ $>$ 17\,\Msun\ according to a Salpeter initial mass function (IMF). However, stochastic sampling of the IMF causes extreme variations in the ratio of B to O-stars \citep{Lamb2013}. Given a single O-star, a surprising 15\% would have no or only one B-star associated with it based on random drawings from the IMF. On the other extreme, an isolated O-star is born with $\ge$\,25 B-stars about 15\% of the time. About 68\% of the time (1$\sigma$ interval), a \textsc{Tip of the Iceberg} O-star is born with $\ge$\,5 B-stars. We therefore designate the 223 \textsc{Field} O-stars (17\%) with $N_{\rm B}$ $\ge$ 5 B-stars within $r_{\rm proj}$~$<$~2~pc as \textsc{Tip of the Iceberg} O-stars. The remaining 1,096 \textsc{Field} O-stars (83\%) are classified as \textsc{Ejected} systems, i.e., both walkaway and runaway O-stars (see Fig.~\ref{fig:flowchart} and Table~\ref{tab:Ostars}). 

In the bottom row of Fig.~\ref{fig:clusters}, we display example 50-pc\,$\times$\,50-pc DSS images of selected \textsc{Tip of the Iceberg} O-stars. The examples are shown left to right in order of increasing B-star density, starting with the minimum threshold of $N_{\rm B}$ $=$ 5 B-stars within $r_{\rm proj}$~$<$~2~pc. We find that 201/223 = 90\% of the \textsc{Tip of the Iceberg} O-stars are located in low-mass clusters with only $N_{\rm B}$~=~5\,-\,14 B-stars within $r_{\rm proj}$~$<$~2~pc (first three examples in Fig.~\ref{fig:clusters}). The remaining 10\% reside in old, dense clusters where all but the least massive O-star has already evolved off of the MS. In the bottom-right image in Fig.~\ref{fig:clusters}, we show an example \textsc{Tip of the Iceberg} O-star that is technically a \textsc{Field} O-star, i.e., there are no other O-stars within 27 pc, but is buried in a dense, old cluster containing dozens of B-dwarfs and several red supergiants. 

We confirm earlier studies that showed that only a small fraction of field O-stars formed in situ \citep{deWit2004,DorigoJones2020,Vargas-Salazar2020}. For example, \citet{deWit2004} found that only 5/43 = 12\%\,$\pm$\,5\% of Galactic field O-stars are tip of the icebergs, consistent with our estimate of 17\%. \citet{Vargas-Salazar2020} measured a somewhat smaller tip of the iceberg fraction of 5\% for their sample of field OB stars in the SMC. The difference might be due to their inclusion of early-B stars in their field sample, whereas we focus strictly on field O-stars with M$_{\rm I}$ $<$ $-$3.5. \citet{Vargas-Salazar2020} used the RIOTS4 sample of SMC field OB stars selected to not have any other OB stars within projected separations $r_{\rm proj}$ $<$ 28~pc \citep{Oey2004,Lamb2016}. Since even early-B stars greatly outnumber the O-stars, they have a strong effect in defining the RIOTS4 field star sample when considering their proximity as an exclusion criterion. Thus, by probing further down the IMF, their OB field sample is more stringent in excluding objects with neighboring OB stars. Compared to our field O-star sample, their field OB stars are much more isolated and less likely to have formed in situ.  The tip-of-the-iceberg stars of \citet{Vargas-Salazar2025} are defined by measured stellar density enhancements that are rarely visually apparent, contrasting significantly with the nature of our \textsc{Tip of the Iceberg} O-stars. In any case, only a small minority of \textsc{Field} O-stars are \textsc{Tip of the Iceberg}. The vast majority of massive field stars are ejected walkaways and runaways.

The EB fraction of \textsc{Tip of the Iceberg} O-stars is 9.0\,$\pm$\,1.9\%, which matches the \textsc{Cluster} value of 9.4\%\,$\pm$\,0.5\%. O-stars born in dense clusters, sparse associations, or even low-mass isolated clusters in the field all exhibit the same EB fraction. We again conclude that the formation process of close massive binaries mostly depends on small-scale gas/disk physics. The stellar density of the surrounding environment does not strongly affect the close binary properties of O-stars that remain in clusters.

In contrast, the EB fraction of \textsc{Ejected} O-stars is 6.2\%\,$\pm$\,0.7\%, which is measurably less than the \textsc{Cluster} value of 9.4\%\,$\pm$\,0.5\%, statistically significant at the 3.7$\sigma$ level. \textsc{Ejected} O-stars have a lower close binary fraction, consistent with previous observations \citep{Gies1986,Gies1987,Mason2009,DorigoJones2020,Vargas-Salazar2025}. For example, \citet{Mason2009} measured spectroscopic binary fractions of 57\% for Galactic O-stars in clusters and associations, 46\% for field O-stars, but only 29\% for Galactic runaway O-stars (see their Table 5). In their sample of 55 field OB stars in the wing of the SMC, \citet{Vargas-Salazar2025} identified three confirmed and two candidate EBs. Their EB fraction of (3\,-\,5)/55 = 5.5\%\,-\,9.1\% encompasses our measurement of 6.2\% for field O-stars in the LMC.

According to numerical models of N-body interactions and dynamical ejections, the binary fraction is predicted to decrease with respect to ejection velocity \citep{Perets2012,Oh2016}. To test this hypothesis, we cross-reference our 1,096 \textsc{Ejected} O-stars with {\it Gaia}~DR3 \citep{GaiaDR3}, where we find 1,089 unique matches within 1'' and with measured proper motions. The median proper motion of our \textsc{Ejected} O-stars is PM$_{\alpha}$ = 1.78~mas~yr$^{-1}$ and PM$_{\delta}$ = 0.11~mas~yr$^{-1}$, consistent with the bulk motion of the LMC \citep{JimenezArranz2023}. However, we notice a slight trend in proper motion with respect to position within the LMC. For each \textsc{Ejected} O-star, we therefore define the local standard of rest as the median proper motion vector of the 30 nearest other \textsc{Ejected} O-stars. The net proper motion PM$_{\rm net}$ is then converted into a projected velocity via the standard relation:

\begin{equation}
v_{\rm proj}({\rm km~s}^{-1}) = 4.74~{\rm PM}_{\rm net} (''/{\rm yr})~d(\rm{pc}) .
\end{equation} 

\begin{figure}[t!]
\centerline{
\includegraphics[trim=0.7cm 0.1cm 0.2cm 0.1cm, clip=true, width=3.3in]{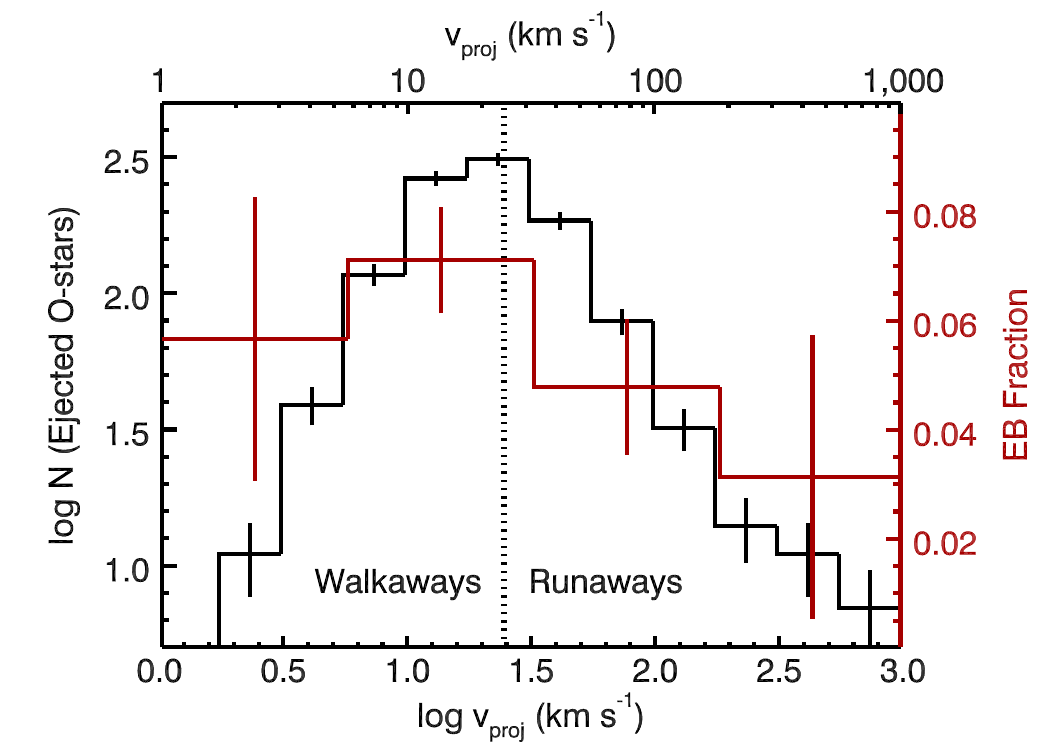}}
\caption{Number of \textsc{Ejected} O-stars as a function of projected velocity (black). We distinguish ejected \textsc{Walkaways} with $v_{\rm proj}$~$<$~24.5 km~s$^{-1}$ from \textsc{Runaway} O-stars with $v_{\rm proj}$~$>$~24.5~km~s$^{-1}$ (dotted black). The EB fraction (red) decreases with velocity, qualitatively consistent with dynamical models of N-body interactions.}
\label{fig:Ejected}
\end{figure}

In Fig.~\ref{fig:Ejected}, we display the distribution of residual projected velocities for our \textsc{Ejected} O-stars. The average projected velocity is $\langle v_{\rm proj} \rangle$~=~22~km~s$^{-1}$. Assuming ejection in isotropic directions, the walkaway/runaway threshold of $v$~=~30~km~s$^{-1}$ corresponds to a projected velocity of v$_{\rm proj}$~=~24.5~km~s$^{-1}$.
We classify the 476 \textsc{Ejected} O-stars (43\%) that exceed $v_{\rm proj}$~$>$~24.5~km~s$^{-1}$ as \textsc{Runaways} (see Fig.~\ref{fig:flowchart}). The remaining 620 \textsc{Ejected} O-stars (57\%) with $v_{\rm proj}$~$<$~24.5~km~s$^{-1}$ are \textsc{Walkaways}, including the seven systems without measured proper motions.

In a magnitude-limited sample of 195 Galactic O-stars, \citet{Gies1987} identified 136 (70\%) in clusters and associations, 43 (22\%) in the field but not runaway, and 16 (8\%) as runaways with peculiar velocities exceeding $v$~$>$~30~km~s$^{-1}$. In the LMC, we find similar occurrence rates for our sample of O-stars: 73\% in \textsc{Clusters}, including associations, 17\% as \textsc{Tip of the Iceberg} or field \textsc{Walkaways}, and the remaining 10\% as field \textsc{Runaways}. The relative proportions of O-stars in different environments are consistent between the Milky Way and LMC. Thus the formation processes of massive stars that produce O-star clusterings and ejected O-stars occur similarly in the two galaxies.

As predicted, the EB fraction decreases with projected velocity among our \textsc{Ejected} O-stars (red histogram in Fig.~\ref{fig:Ejected}). The EB fractions of the \textsc{Walkaway} and \textsc{Runaway} O-stars are 7.3\%\,$\pm$\,1.0\% and 4.7\%\,$\pm$\,1.0\%, respectively (Fig.~\ref{fig:flowchart} and Table~\ref{tab:Ostars}). The ratio in the EB fractions between \textsc{Runaway} and \textsc{Cluster} O-stars is 4.7\%/9.4\% = 50\%. In other words, the close binary fraction of \textsc{Runaway} O-stars is reduced to half of their \textsc{Cluster} counterparts.  \citet{Mason2009} measured a ratio in the spectroscopic binary fractions between Galactic runaway and cluster O-stars of 29\%/57\% = 51\%, similar to our ratio for EBs.

The EB fraction of \textsc {Walkaway} O-stars in the field is reduced to 7.3\%/9.4\% = 78\% of the \textsc{Cluster} value. If most field walkaways derived from supernova kicks in binaries, then we would expect the \textsc{Walkaway} EB fraction to be considerably lower. The majority of \textsc{Ejected} O-stars in the field therefore derive from N-body dynamical interactions, consistent with previous findings \citep{DorigoJones2020,Phillips2024}. 

Most O-star companions that are kicked by supernova explosions in binaries remain in the vicinity of their birth clusters, leading to the lower EB fraction observed in \textsc{Old Clusters} (see Section~\ref{sec:cluster}). Only a small fraction escape to $r_{\rm proj}$ $>$ 27~pc from other O-stars to become field walkaways or field runaways. For example, dozens of massive runaways have been detected near the R136 cluster / 30 Doradus within the LMC \citep{Sana2022,Stoop2024}. Although the runaways surround the periphery of 30 Doradus, a significant majority are located within $r_{\rm proj}$ $<$ 27~pc of at least one other massive star and therefore are not in the \textsc{Field}. After long-term radial velocity monitoring, \citet{Sana2022} identified only one spectroscopic binary in their sample of 24 massive runaways. The overabundance of single stars suggests that the majority of non-field massive runaways in the periphery of 30 Doradus were kicked via supernova explosions in binaries. \citet{Sana2022} also identified two populations of runaways: slowly moving but rapidly rotating stars that likely derived from the binary evolution scenario and rapidly moving but slowly rotating stars that likely derived from N-body dynamical ejections. The systematically slower velocities of runaways that derive from supernova kicks in binaries are consistent with numerical models \citep{Renzo2019}.

In the SMC, \citet{DorigoJones2020} estimated that for every OB companion that reaches the field via supernova kicks, there are 1.4 OB companions that remain within projected separations $r_{\rm proj}$~$<$~28~pc of another OB star, i.e., a correction factor of 2.4. The correction factor is likely higher for O-star companions due to two effects. First, O-star companions are systematically more massive and have more inertia. Thus O-star companions are kicked by supernova explosions with slower velocities. According to the binary evolution models of \citet{Renzo2019}, the ratio of walkaways to runaways is 13 for OB companions ($M$ $>$ 7.5\,\Msun) but 20 for O-star companions ($M$ $>$ 15\,\Msun). Second, ejected O-star companions have shorter lifetimes than ejected early-B companions. For a pair of O-stars, the timespan between the first O-star exploding as a supernova and the companion O-star evolving off the MS is considerably shorter than for its O+B binary counterpart. O-star companions ejected via supernova kicks have slower velocities and shorter lifetimes, and therefore they are less likely to escape into the field. 

In the SMC, 25\%\,-\,30\% of ejected OB stars derive from supernova kicks in binaries \citep{DorigoJones2020,Phillips2024}. These rates increase if we consider the two-step process whereby massive binaries are initially ejected from their birth clusters via N-body dynamical interactions and subsequently the OB companions receive an extra kick as the primaries undergo a supernova explosion. Including two-steppers, then 30\%\,-\,45\% of ejected OB stars in the SMC derive from the binary supernova scenario (see Table~4 in \citealt{Phillips2024}). Although two-step ejections will readily escape into the field environment, only 1/2.4 = 42\% of OB companions from the pure supernova channel will reach the field \citep{DorigoJones2020}.  Thus only 20\%\,-\,35\% of OB stars in the field of the SMC are ejected via supernova explosions, including two-steppers. 

For our LMC sample, the high EB fraction of field \textsc{Walkaway} O-stars dictates that at most $<$\,22\% derive from supernova kicks in binaries. Similarly, the EB fraction of \textsc{Ejected} O-stars is reduced to 6.2\%/9.4\% = 66\% of the \textsc{Cluser} value, and thus at most $<$\,34\% of \textsc{Ejected} O-stars are kicked by supernova explosions in binaries, including two-steppers. Considering the small contribution from \textsc{Tip of the Iceberg} O-stars, then at most $<$\,28\% of \textsc{Field} O-stars are ejected via supernova explosions. Our upper limit is similar to the measurements of \citet{DorigoJones2020} and \citet{Phillips2024}. Any slight quantitative difference stems from the different selection criteria where we focus strictly on O-stars while the cited studies included early-B stars.  Compared to early-B companions, O-star companions kicked by supernova explosions are less likely to have reached the field (see above). In any case, the majority of ejected OB stars in the SMC field and ejected O-stars in the LMC field derive from N-body dynamical interactions. Only a minority derive from supernova kicks, including two-steppers. 

Field \textsc{Walkaway} and especially field \textsc{Runaway} O-stars derive mostly from N-body dynamical ejections. According to the most realistic N-body dynamical model in \citet[][their~model~009]{Perets2012}, the binary fraction decreases from 40\% for walkaways with $v$ $<$ 30~km~s$^{-1}$ to 25\% for typical runaways across $v$ = 45\,-\,70\,km~s$^{-1}$ (blue histogram in their Fig.~2). The reduction in the binary fraction between their runaways versus walkaways is 25\%/40\% = 62\%. For our \textsc{Ejected} O-stars, the measured reduction in the EB fraction between \textsc{Runaways} versus \textsc{Walkaways} is 4.7\%/7.3\% = 64\%, consistent with the N-body model predictions.  We re-iterate that the supernova kick scenario may dominate for non-field massive runaways that remain in the vicinity of their birth cluster \citep[][see above]{Sana2022,Stoop2024}. For walkways and runaways in the \textsc{Field}, however, our EB measurements demonstrate that most come from N-body dynamical ejections.

\begin{figure}[t!]
\centerline{
\includegraphics[trim=0.7cm 0.1cm 0.1cm 0.1cm, clip=true, width=3.4in]{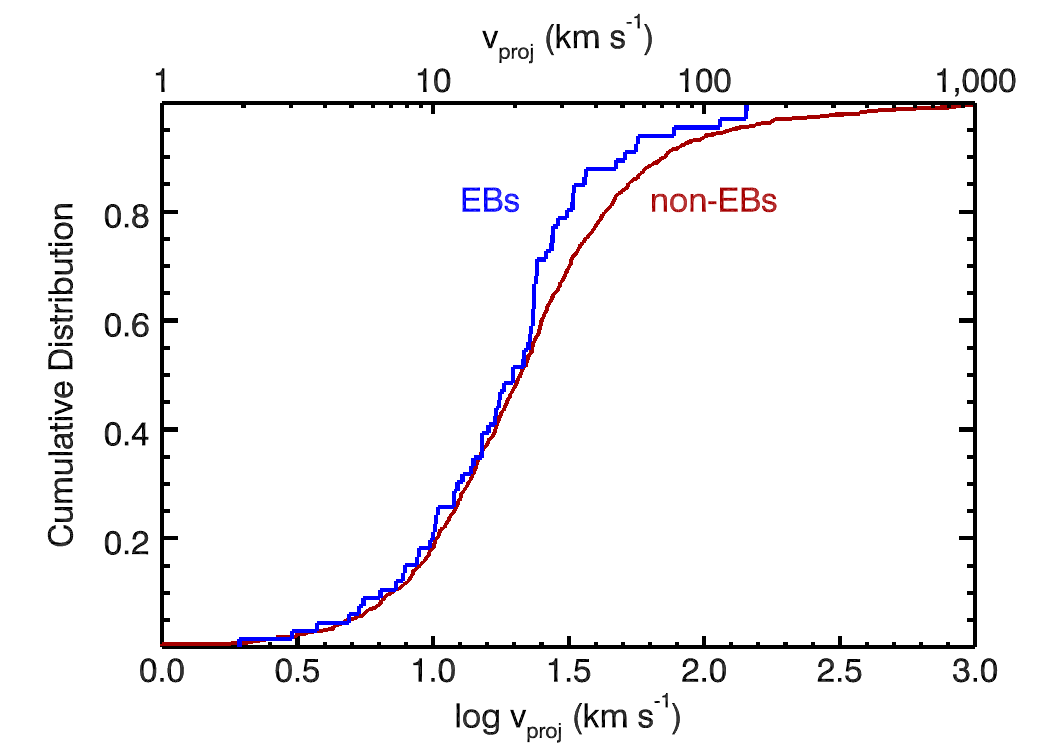}}
\caption{Cumulative projected velocity distribution of \textsc{Ejected} O-stars separated into EBs (blue) and non-EBs (red). There is a statistically significant deficit of high-velocity EBs above $v$~$>$~20~km~s$^{-1}$ compared to non-EBs.}
\label{fig:Cumv}
\end{figure}

To further illustrate this trend, we display in Fig.~\ref{fig:Cumv} the cumulative distribution of projected velocities of our \textsc{Ejected} O-stars separated into EBs and non-EBs. The two distributions are similar below $v_{\rm proj}$~$<$~20~km~s$^{-1}$, but at higher velocities the EBs exhibit a noticeable deficit. According to an Anderson-Darling test, the EB and non-EB distributions are inconsistent with each other at the 2.5$\sigma$ significance level. Faster \textsc{Ejected} O-stars are less likely to host close binaries.

The median orbital period of the fastest \textsc{Runaways} with $v_{\rm proj}$~$>$~30 km s$^{-1}$ is $P$ = 2.76\,$\pm$\,0.27 days, which is less than the \textsc{Overall} value, statistically significant at the 2.6$\sigma$ level. N-body interactions that launch fast runaways tend to preferentially  disrupt wider binaries with $v_{\rm orb}$ $\lesssim$ $v$ \citep{Perets2012}. Runaway EBs have a larger deficit of wider companions, consistent with the measured shift toward shorter periods.

\section{Summary}
\label{sec:summary}

We summarize our main results as follows:

\begin{itemize}
\item After correcting for geometrical eclipse probabilities, our LMC O-type binaries follow a power-law period distribution $f_{\rm logP}$~$\propto$~(log\,$P$)$^\Pi$ with $\Pi$ = $-$0.34\,$\pm$\,0.06 across $P$~=~2.5\,-\,200~days. The observed distribution is skewed towards shorter periods compared to Opik's law ($\Pi$~=~0) but is slightly shallower than the slope of $\Pi$~=~$-$0.55\,$\pm$\,0.22 measured for Galactic O-type spectroscopic binaries. 

\item The EB fraction of O-stars in LMC \textsc{Young Clusters} is 10.2\%\,$\pm$\,0.6\%, which matches the 10.8\%\,$\pm$\,2.1\% for their counterparts in Milky Way young clusters. The slight metallicity difference between Galactic O-stars ([Fe/H] = 0.0) and LMC O-stars ([Fe/H] = $-$0.4) does not affect their close binary properties. 

\item The EB fraction of O-stars in \textsc{Old Clusters} is 5.5\,$\pm$\,0.9\%, which is about half the value observed in \textsc{Young Clusters}, statistically discrepant at the 4.3$\sigma$ level. The close binary fraction of older O-stars is reduced due to the effects of binary evolution, including mergers and the original primaries evolving into compact remnants. Most O-star companions that are kicked by supernova explosions in binaries remain in the vicinity of their birth clusters. At most $<$\,28\% of \textsc{Field} O-stars are secondaries that were kicked when their primaries exploded as supernovae.

\item The EB fractions of O-stars in \textsc{Young Associations}, \textsc{Young Average Clusters}, and \textsc{Young Dense Clusters} are consistent with each other at 10.6\%\,$\pm$\,1.1\%, 9.2\%\,$\pm$\,0.8\%, and 11.3\%\,$\pm$\,1.1\%, respectively. This suggests that the formation process of close massive binaries depends on small-scale gas physics, e.g., fragmentation and migration within protostellar disks. N-body interactions that scale with cluster density do not significantly affect the close binary properties of massive stars in clusters.

\item The proportions of LMC \textsc{Cluster} O-stars (73\%),  \textsc{Field} non-runaways (17\%), and \textsc{Runaways} (10\%) match the ratios measured for Galactic O-stars (70\%:22\%:8\%). Clustering of O-stars and their ejections into the field occur similarly in the LMC and Milky Way. 

\item About 17\% of \textsc{Field} O-stars exhibit an overdensity of surrounding B-stars, demonstrating that they are \textsc{Tip of the Iceberg} O-stars that formed in situ in the field within isolated, low-mass clusters. We caution that comparison of this value to that found in other studies should account for the varying selection criteria and definition of field O-stars. The \textsc{Tip of the Iceberg} EB fraction of 9.0\%\,$\pm$\,1.9\% matches the cluster value, again suggesting that cluster density does not affect the properties of close massive binaries that remain in clusters.

\item The remaining 83\% of \textsc{Field} O-stars were ejected from their birth clusters, a majority ($>$66\%) of which were ejected via N-body dynamical interactions. The EB fractions of field \textsc{Walkaway} ($v_{\rm proj}$~$<$~24.5~km~s$^{-1}$) and field \textsc{Runaway} ($v_{\rm proj}$~$>$~24.5~km~s$^{-1}$) O-stars are 7.3\%\,$\pm$\,1.0\% and 4.7\%\,$\pm$\,1.0\%, respectively, qualitatively consistent with predictions from N-body dynamical models.

\end{itemize}

\bibliographystyle{aasjournalv7}                       
\bibliography{moe_biblio}

\begin{thebibliography}{}
\expandafter\ifx\csname natexlab\endcsname\relax\def\natexlab#1{#1}\fi
\providecommand{\url}[1]{\href{#1}{#1}}
\providecommand{\dodoi}[1]{doi:~\href{http://doi.org/#1}{\nolinkurl{#1}}}
\providecommand{\doeprint}[1]{\href{http://ascl.net/#1}{\nolinkurl{http://ascl.net/#1}}}
\providecommand{\doarXiv}[1]{\href{https://arxiv.org/abs/#1}{\nolinkurl{https://arxiv.org/abs/#1}}}

\bibitem[{L.~A. {Almeida} {et~al.}(2017){Almeida}, {Sana}, {Taylor}, {Barb{\'a}}, {Bonanos}, {Crowther}, {Damineli}, {de Koter}, {de Mink}, {Evans}, {Gieles}, {Grin}, {H{\'e}nault-Brunet}, {Langer}, {Lennon}, {Lockwood}, {Ma{\'\i}z Apell{\'a}niz}, {Moffat}, {Neijssel}, {Norman}, {Ram{\'\i}rez-Agudelo}, {Richardson}, {Schootemeijer}, {Shenar}, {Soszy{\'n}ski}, {Tramper}, \& {Vink}}]{Almeida2017}
{Almeida}, L.~A., {Sana}, H., {Taylor}, W., {et~al.} 2017, \bibinfo{title}{{The Tarantula Massive Binary Monitoring. I. Observational campaign and OB-type spectroscopic binaries},} \aap, 598, A84, \dodoi{10.1051/0004-6361/201629844}

\bibitem[{P. {Artymowicz} {et~al.}(1991){Artymowicz}, {Clarke}, {Lubow}, \& {Pringle}}]{Artymowicz1991}
{Artymowicz}, P., {Clarke}, C.~J., {Lubow}, S.~H., \& {Pringle}, J.~E. 1991, \bibinfo{title}{{The Effect of an External Disk on the Orbital Elements of a Central Binary},} \apjl, 370, L35, \dodoi{10.1086/185971}

\bibitem[{M.~R. {Bate}(1998){Bate}}]{Bate1998}
{Bate}, M.~R. 1998, \bibinfo{title}{{Collapse of a Molecular Cloud Core to Stellar Densities: The First Three-dimensional Calculations},} \apjl, 508, L95, \dodoi{10.1086/311719}

\bibitem[{M.~R. {Bate}(2009){Bate}}]{Bate2009}
{Bate}, M.~R. 2009, \bibinfo{title}{{Stellar, brown dwarf and multiple star properties from hydrodynamical simulations of star cluster formation},} \mnras, 392, 590, \dodoi{10.1111/j.1365-2966.2008.14106.x}

\bibitem[{M.~R. {Bate} \& I.~A. {Bonnell}(1997){Bate} \& {Bonnell}}]{Bate1997}
{Bate}, M.~R., \& {Bonnell}, I.~A. 1997, \bibinfo{title}{{Accretion during binary star formation - II. Gaseous accretion and disc formation},} \mnras, 285, 33, \dodoi{10.1093/mnras/285.1.33}

\bibitem[{M.~R. {Bate} {et~al.}(2002){Bate}, {Bonnell}, \& {Bromm}}]{Bate2002}
{Bate}, M.~R., {Bonnell}, I.~A., \& {Bromm}, V. 2002, \bibinfo{title}{{The formation of close binary systems by dynamical interactions and orbital decay},} \mnras, 336, 705, \dodoi{10.1046/j.1365-8711.2002.05775.x}

\bibitem[{M.~R. {Bate} {et~al.}(1995){Bate}, {Bonnell}, \& {Price}}]{Bate1995}
{Bate}, M.~R., {Bonnell}, I.~A., \& {Price}, N.~M. 1995, \bibinfo{title}{{Modelling accretion in protobinary systems},} \mnras, 277, 362, \dodoi{10.1093/mnras/277.2.362}

\bibitem[{P. {Battinelli}(1991){Battinelli}}]{Battinelli1991}
{Battinelli}, P. 1991, \bibinfo{title}{{A new identification technique for OB associations : OB associations in the Small Magellanic Cloud.},} \aap, 244, 69

\bibitem[{A. {Blaauw}(1961){Blaauw}}]{Blaauw1961}
{Blaauw}, A. 1961, \bibinfo{title}{{On the origin of the O- and B-type stars with high velocities (the ``run-away'' stars), and some related problems},} \bain, 15, 265

\bibitem[{A.~P. {Boss}(1986){Boss}}]{Boss1986}
{Boss}, A.~P. 1986, \bibinfo{title}{{Protostellar Formation in Rotating Interstellar Clouds. V. Nonisothermal Collapse and Fragmentation},} \apjs, 62, 519, \dodoi{10.1086/191150}

\bibitem[{S.~E. {de~Mink} {et~al.}(2014){de~Mink}, {Sana}, {Langer}, {Izzard}, \& {Schneider}}]{deMink2014}
{de~Mink}, S.~E., {Sana}, H., {Langer}, N., {Izzard}, R.~G., \& {Schneider}, F.~R.~N. 2014, \bibinfo{title}{{The Incidence of Stellar Mergers and Mass Gainers among Massive Stars},} \apj, 782, 7, \dodoi{10.1088/0004-637X/782/1/7}

\bibitem[{W.~J. {de~Wit} {et~al.}(2004){de~Wit}, {Testi}, {Palla}, {Vanzi}, \& {Zinnecker}}]{deWit2004}
{de~Wit}, W.~J., {Testi}, L., {Palla}, F., {Vanzi}, L., \& {Zinnecker}, H. 2004, \bibinfo{title}{{The origin of massive O-type field stars. I. A search for clusters},} \aap, 425, 937, \dodoi{10.1051/0004-6361:20040454}

\bibitem[{W.~J. {de Wit} {et~al.}(2005){de Wit}, {Testi}, {Palla}, \& {Zinnecker}}]{deWit2005}
{de Wit}, W.~J., {Testi}, L., {Palla}, F., \& {Zinnecker}, H. 2005, \bibinfo{title}{{The origin of massive O-type field stars: II. Field O stars as runaways},} \aap, 437, 247, \dodoi{10.1051/0004-6361:20042489}

\bibitem[{J. {Dorigo Jones} {et~al.}(2020){Dorigo Jones}, {Oey}, {Paggeot}, {Castro}, \& {Moe}}]{DorigoJones2020}
{Dorigo Jones}, J., {Oey}, M.~S., {Paggeot}, K., {Castro}, N., \& {Moe}, M. 2020, \bibinfo{title}{{Runaway OB Stars in the Small Magellanic Cloud: Dynamical versus Supernova Ejections},} \apj, 903, 43, \dodoi{10.3847/1538-4357/abbc6b}

\bibitem[{P.~R. {Dunstall} {et~al.}(2015){Dunstall}, {Dufton}, {Sana}, {Evans}, {Howarth}, {Sim{\'o}n-D{\'\i}az}, {de Mink}, {Langer}, {Ma{\'\i}z Apell{\'a}niz}, \& {Taylor}}]{Dunstall2015}
{Dunstall}, P.~R., {Dufton}, P.~L., {Sana}, H., {et~al.} 2015, \bibinfo{title}{{The VLT-FLAMES Tarantula Survey. XXII. Multiplicity properties of the B-type stars},} \aap, 580, A93, \dodoi{10.1051/0004-6361/201526192}

\bibitem[{ {Gaia~Collaboration et al.}(2022){Gaia~Collaboration et al.}}]{GaiaDR3}
{Gaia~Collaboration et al.} 2022, \bibinfo{title}{{Gaia Data Release 3: Summary of the content and survey properties},} arXiv e-prints, arXiv:2208.00211, \dodoi{10.48550/arXiv.2208.00211}

\bibitem[{D.~R. {Gies}(1987){Gies}}]{Gies1987}
{Gies}, D.~R. 1987, \bibinfo{title}{{The Kinematical and Binary Properties of Association and Field O Stars},} \apjs, 64, 545, \dodoi{10.1086/191208}

\bibitem[{D.~R. {Gies} \& C.~T. {Bolton}(1986){Gies} \& {Bolton}}]{Gies1986}
{Gies}, D.~R., \& {Bolton}, C.~T. 1986, \bibinfo{title}{{The Binary Frequency and Origin of the OB Runaway Stars},} \apjs, 61, 419, \dodoi{10.1086/191118}

\bibitem[{D. {Graczyk} {et~al.}(2011){Graczyk}, {Soszy{\'n}ski}, {Poleski}, {Pietrzy{\'n}ski}, {Udalski}, {Szyma{\'n}ski}, {Kubiak}, {Wyrzykowski}, \& {Ulaczyk}}]{Graczyk2011}
{Graczyk}, D., {Soszy{\'n}ski}, I., {Poleski}, R., {et~al.} 2011, \bibinfo{title}{{The Optical Gravitational Lensing Experiment. The OGLE-III Catalog of Variable Stars. XII. Eclipsing Binary Stars in the Large Magellanic Cloud},} ACTAA, 61, 103, \dodoi{10.48550/arXiv.1108.0446}

\bibitem[{D.~C. {Heggie}(1975){Heggie}}]{Heggie1975}
{Heggie}, D.~C. 1975, \bibinfo{title}{{Binary evolution in stellar dynamics.},} \mnras, 173, 729, \dodoi{10.1093/mnras/173.3.729}

\bibitem[{R. {Hoogerwerf} {et~al.}(2001){Hoogerwerf}, {de Bruijne}, \& {de Zeeuw}}]{Hoogerwerf2001}
{Hoogerwerf}, R., {de Bruijne}, J.~H.~J., \& {de Zeeuw}, P.~T. 2001, \bibinfo{title}{{On the origin of the O and B-type stars with high velocities. II. Runaway stars and pulsars ejected from the nearby young stellar groups},} \aap, 365, 49, \dodoi{10.1051/0004-6361:20000014}

\bibitem[{{\'O}. {Jim{\'e}nez-Arranz} {et~al.}(2023){Jim{\'e}nez-Arranz}, {Romero-G{\'o}mez}, {Luri}, {McMillan}, {Antoja}, {Chemin}, {Roca-F{\`a}brega}, {Masana}, \& {Muros}}]{JimenezArranz2023}
{Jim{\'e}nez-Arranz}, {\'O}., {Romero-G{\'o}mez}, M., {Luri}, X., {et~al.} 2023, \bibinfo{title}{{Kinematic analysis of the Large Magellanic Cloud using Gaia DR3},} \aap, 669, A91, \dodoi{10.1051/0004-6361/202244601}

\bibitem[{D.~C. {Kiminki} \& H.~A. {Kobulnicky}(2012){Kiminki} \& {Kobulnicky}}]{Kiminki2012}
{Kiminki}, D.~C., \& {Kobulnicky}, H.~A. 2012, \bibinfo{title}{{An Updated Look at Binary Characteristics of Massive Stars in the Cygnus OB2 Association},} \apj, 751, 4, \dodoi{10.1088/0004-637X/751/1/4}

\bibitem[{L.~G. {Kiseleva} {et~al.}(1998){Kiseleva}, {Eggleton}, \& {Mikkola}}]{Kiseleva1998}
{Kiseleva}, L.~G., {Eggleton}, P.~P., \& {Mikkola}, S. 1998, \bibinfo{title}{{Tidal friction in triple stars},} \mnras, 300, 292, \dodoi{10.1046/j.1365-8711.1998.01903.x}

\bibitem[{K.~M. {Kratter} \& C.~D. {Matzner}(2006){Kratter} \& {Matzner}}]{Kratter2006}
{Kratter}, K.~M., \& {Matzner}, C.~D. 2006, \bibinfo{title}{{Fragmentation of massive protostellar discs},} \mnras, 373, 1563, \dodoi{10.1111/j.1365-2966.2006.11103.x}

\bibitem[{P. {Kroupa}(1995){Kroupa}}]{Kroupa1995}
{Kroupa}, P. 1995, \bibinfo{title}{{Inverse dynamical population synthesis and star formation},} \mnras, 277, 1491, \dodoi{10.1093/mnras/277.4.1491}

\bibitem[{J.~B. {Lamb} {et~al.}(2013){Lamb}, {Oey}, {Graus}, {Adams}, \& {Segura-Cox}}]{Lamb2013}
{Lamb}, J.~B., {Oey}, M.~S., {Graus}, A.~S., {Adams}, F.~C., \& {Segura-Cox}, D.~M. 2013, \bibinfo{title}{{The Initial Mass Function of Field OB Stars in the Small Magellanic Cloud},} \apj, 763, 101, \dodoi{10.1088/0004-637X/763/2/101}

\bibitem[{J.~B. {Lamb} {et~al.}(2016){Lamb}, {Oey}, {Segura-Cox}, {Graus}, {Kiminki}, {Golden-Marx}, \& {Parker}}]{Lamb2016}
{Lamb}, J.~B., {Oey}, M.~S., {Segura-Cox}, D.~M., {et~al.} 2016, \bibinfo{title}{{The Runaways and Isolated O-Type Star Spectroscopic Survey of the SMC (RIOTS4)},} \apj, 817, 113, \dodoi{10.3847/0004-637X/817/2/113}

\bibitem[{B.~D. {Mason} {et~al.}(2009){Mason}, {Hartkopf}, {Gies}, {Henry}, \& {Helsel}}]{Mason2009}
{Mason}, B.~D., {Hartkopf}, W.~I., {Gies}, D.~R., {Henry}, T.~J., \& {Helsel}, J.~W. 2009, \bibinfo{title}{{The High Angular Resolution Multiplicity of Massive Stars},} \aj, 137, 3358, \dodoi{10.1088/0004-6256/137/2/3358}

\bibitem[{M. {Moe} \& R. {Di~Stefano}(2013){Moe} \& {Di~Stefano}}]{Moe2013}
{Moe}, M., \& {Di~Stefano}, R. 2013, \bibinfo{title}{{The Close Binary Properties of Massive Stars in the Milky Way and Low-metallicity Magellanic Clouds},} \apj, 778, 95, \dodoi{10.1088/0004-637X/778/2/95}

\bibitem[{M. {Moe} \& R. {Di Stefano}(2015){Moe} \& {Di Stefano}}]{Moe2015a}
{Moe}, M., \& {Di Stefano}, R. 2015, \bibinfo{title}{{A New Class of Nascent Eclipsing Binaries with Extreme Mass Ratios},} \apj, 801, 113, \dodoi{10.1088/0004-637X/801/2/113}

\bibitem[{M. {Moe} \& R. {Di~Stefano}(2017){Moe} \& {Di~Stefano}}]{Moe2017}
{Moe}, M., \& {Di~Stefano}, R. 2017, \bibinfo{title}{{Mind Your Ps and Qs: The Interrelation between Period (P) and Mass-ratio (Q) Distributions of Binary Stars},} \apjs, 230, 15, \dodoi{10.3847/1538-4365/aa6fb6}

\bibitem[{M. {Moe} \& K.~M. {Kratter}(2018){Moe} \& {Kratter}}]{Moe2018}
{Moe}, M., \& {Kratter}, K.~M. 2018, \bibinfo{title}{{Dynamical Formation of Close Binaries during the Pre-main-sequence Phase},} \apj, 854, 44, \dodoi{10.3847/1538-4357/aaa6d2}

\bibitem[{S. {Naoz} \& D.~C. {Fabrycky}(2014){Naoz} \& {Fabrycky}}]{Naoz2014}
{Naoz}, S., \& {Fabrycky}, D.~C. 2014, \bibinfo{title}{{Mergers and Obliquities in Stellar Triples},} \apj, 793, 137, \dodoi{10.1088/0004-637X/793/2/137}

\bibitem[{P.~K. {Nayak} {et~al.}(2016){Nayak}, {Subramaniam}, {Choudhury}, {Indu}, \& {Sagar}}]{Nayak2016}
{Nayak}, P.~K., {Subramaniam}, A., {Choudhury}, S., {Indu}, G., \& {Sagar}, R. 2016, \bibinfo{title}{{Star clusters in the Magellanic Clouds - I. Parametrization and classification of 1072 clusters in the LMC},} \mnras, 463, 1446, \dodoi{10.1093/mnras/stw2043}

\bibitem[{M.~S. {Oey} {et~al.}(2004){Oey}, {King}, \& {Parker}}]{Oey2004}
{Oey}, M.~S., {King}, N.~L., \& {Parker}, J.~W. 2004, \bibinfo{title}{{Massive Field Stars and the Stellar Clustering Law},} \aj, 127, 1632, \dodoi{10.1086/381926}

\bibitem[{S.~S.~R. {Offner} {et~al.}(2023){Offner}, {Moe}, {Kratter}, {Sadavoy}, {Jensen}, \& {Tobin}}]{Offner2023}
{Offner}, S.~S.~R., {Moe}, M., {Kratter}, K.~M., {et~al.} 2023, \bibinfo{title}{{The Origin and Evolution of Multiple Star Systems},} in Astronomical Society of the Pacific Conference Series, Vol. 534, Protostars and Planets VII, ed. S.~{Inutsuka}, Y.~{Aikawa}, T.~{Muto}, K.~{Tomida}, \& M.~{Tamura}, 275, \dodoi{10.48550/arXiv.2203.10066}

\bibitem[{S. {Oh} \& P. {Kroupa}(2016){Oh} \& {Kroupa}}]{Oh2016}
{Oh}, S., \& {Kroupa}, P. 2016, \bibinfo{title}{{Dynamical ejections of massive stars from young star clusters under diverse initial conditions},} \aap, 590, A107, \dodoi{10.1051/0004-6361/201628233}

\bibitem[{M.~J. {Pecaut} \& E.~E. {Mamajek}(2013){Pecaut} \& {Mamajek}}]{Pecaut2013}
{Pecaut}, M.~J., \& {Mamajek}, E.~E. 2013, \bibinfo{title}{{Intrinsic Colors, Temperatures, and Bolometric Corrections of Pre-main-sequence Stars},} \apjs, 208, 9, \dodoi{10.1088/0067-0049/208/1/9}

\bibitem[{H.~B. {Perets} \& L. {{\v{S}}ubr}(2012){Perets} \& {{\v{S}}ubr}}]{Perets2012}
{Perets}, H.~B., \& {{\v{S}}ubr}, L. 2012, \bibinfo{title}{{The Properties of Dynamically Ejected Runaway and Hyper-runaway Stars},} \apj, 751, 133, \dodoi{10.1088/0004-637X/751/2/133}

\bibitem[{G.~D. {Phillips} {et~al.}(2024){Phillips}, {Oey}, {Cuevas}, {Castro}, \& {Kothari}}]{Phillips2024}
{Phillips}, G.~D., {Oey}, M.~S., {Cuevas}, M., {Castro}, N., \& {Kothari}, R. 2024, \bibinfo{title}{{Runaway OB Stars in the Small Magellanic Cloud. III. Updated Kinematics and Insights into Dynamical versus Supernova Ejections},} \apj, 966, 243, \dodoi{10.3847/1538-4357/ad3909}

\bibitem[{G. {Pietrzy{\'n}ski} {et~al.}(2019){Pietrzy{\'n}ski}, {Graczyk}, {Gallenne}, {Gieren}, {Thompson}, {Pilecki}, {Karczmarek}, {G{\'o}rski}, {Suchomska}, {Taormina}, {Zgirski}, {Wielg{\'o}rski}, {Ko{\l}aczkowski}, {Konorski}, {Villanova}, {Nardetto}, {Kervella}, {Bresolin}, {Kudritzki}, {Storm}, {Smolec}, \& {Narloch}}]{Pietrzynski2019}
{Pietrzy{\'n}ski}, G., {Graczyk}, D., {Gallenne}, A., {et~al.} 2019, \bibinfo{title}{{A distance to the Large Magellanic Cloud that is precise to one per cent},} \nat, 567, 200, \dodoi{10.1038/s41586-019-0999-4}

\bibitem[{M.~C. {Ram{\'\i}rez-Tannus} {et~al.}(2021){Ram{\'\i}rez-Tannus}, {Backs}, {de Koter}, {Sana}, {Beuther}, {Bik}, {Brandner}, {Kaper}, {Linz}, {Henning}, \& {Poorta}}]{RamirezTannus2021}
{Ram{\'\i}rez-Tannus}, M.~C., {Backs}, F., {de Koter}, A., {et~al.} 2021, \bibinfo{title}{{A relation between the radial velocity dispersion of young clusters and their age. Evidence for hardening as the formation scenario of massive close binaries},} \aap, 645, L10, \dodoi{10.1051/0004-6361/202039673}

\bibitem[{M. {Renzo} {et~al.}(2019){Renzo}, {Zapartas}, {de Mink}, {G{\"o}tberg}, {Justham}, {Farmer}, {Izzard}, {Toonen}, \& {Sana}}]{Renzo2019}
{Renzo}, M., {Zapartas}, E., {de Mink}, S.~E., {et~al.} 2019, \bibinfo{title}{{Massive runaway and walkaway stars. A study of the kinematical imprints of the physical processes governing the evolution and explosion of their binary progenitors},} \aap, 624, A66, \dodoi{10.1051/0004-6361/201833297}

\bibitem[{H. {Sana} {et~al.}(2012){Sana}, {de Mink}, {de Koter}, {Langer}, {Evans}, {Gieles}, {Gosset}, {Izzard}, {Le Bouquin}, \& {Schneider}}]{Sana2012}
{Sana}, H., {de Mink}, S.~E., {de Koter}, A., {et~al.} 2012, \bibinfo{title}{{Binary Interaction Dominates the Evolution of Massive Stars},} Science, 337, 444, \dodoi{10.1126/science.1223344}

\bibitem[{H. {Sana} {et~al.}(2013){Sana}, {de Koter}, {de Mink}, {Dunstall}, {Evans}, {H{\'e}nault-Brunet}, {Ma{\'\i}z Apell{\'a}niz}, {Ram{\'\i}rez-Agudelo}, {Taylor}, {Walborn}, {Clark}, {Crowther}, {Herrero}, {Gieles}, {Langer}, {Lennon}, \& {Vink}}]{Sana2013}
{Sana}, H., {de Koter}, A., {de Mink}, S.~E., {et~al.} 2013, \bibinfo{title}{{The VLT-FLAMES Tarantula Survey. VIII. Multiplicity properties of the O-type star population},} \aap, 550, A107, \dodoi{10.1051/0004-6361/201219621}

\bibitem[{H. {Sana} {et~al.}(2022){Sana}, {Ram{\'\i}rez-Agudelo}, {H{\'e}nault-Brunet}, {Mahy}, {Almeida}, {de Koter}, {Bestenlehner}, {Evans}, {Langer}, {Schneider}, {Crowther}, {de Mink}, {Herrero}, {Lennon}, {Gieles}, {Ma{\'\i}z Apell{\'a}niz}, {Renzo}, {Sabbi}, {van Loon}, \& {Vink}}]{Sana2022}
{Sana}, H., {Ram{\'\i}rez-Agudelo}, O.~H., {H{\'e}nault-Brunet}, V., {et~al.} 2022, \bibinfo{title}{{The VLT-FLAMES Tarantula Survey. Observational evidence for two distinct populations of massive runaway stars in 30 Doradus},} \aap, 668, L5, \dodoi{10.1051/0004-6361/202244677}

\bibitem[{F.~R.~N. {Schneider} {et~al.}(2014){Schneider}, {Izzard}, {de Mink}, {Langer}, {Stolte}, {de Koter}, {Gvaramadze}, {Hu{\ss}mann}, {Liermann}, \& {Sana}}]{Schneider2014}
{Schneider}, F.~R.~N., {Izzard}, R.~G., {de Mink}, S.~E., {et~al.} 2014, \bibinfo{title}{{Ages of Young Star Clusters, Massive Blue Stragglers, and the Upper Mass Limit of Stars: Analyzing Age-dependent Stellar Mass Functions},} \apj, 780, 117, \dodoi{10.1088/0004-637X/780/2/117}

\bibitem[{R.~C. {Stone}(1991){Stone}}]{Stone1991}
{Stone}, R.~C. 1991, \bibinfo{title}{{The Space Frequency and Origin of the Runaway O and B Stars},} \aj, 102, 333, \dodoi{10.1086/115880}

\bibitem[{M. {Stoop} {et~al.}(2024){Stoop}, {de Koter}, {Kaper}, {Brands}, {Portegies Zwart}, {Sana}, {Stoppa}, {Gieles}, {Mahy}, {Shenar}, {Guo}, {Nelemans}, \& {Rieder}}]{Stoop2024}
{Stoop}, M., {de Koter}, A., {Kaper}, L., {et~al.} 2024, \bibinfo{title}{{Two waves of massive stars running away from the young cluster R136},} \nat, 634, 809, \dodoi{10.1038/s41586-024-08013-8}

\bibitem[{A. {Tokovinin} \& M. {Moe}(2020){Tokovinin} \& {Moe}}]{Tokovinin2020}
{Tokovinin}, A., \& {Moe}, M. 2020, \bibinfo{title}{{Formation of close binaries by disc fragmentation and migration, and its statistical modelling},} \mnras, 491, 5158, \dodoi{10.1093/mnras/stz3299}

\bibitem[{A. {Udalski} {et~al.}(2008){Udalski}, {Soszynski}, {Szymanski}, {Kubiak}, {Pietrzynski}, {Wyrzykowski}, {Szewczyk}, {Ulaczyk}, \& {Poleski}}]{Udalski2008}
{Udalski}, A., {Soszynski}, I., {Szymanski}, M.~K., {et~al.} 2008, \bibinfo{title}{{The Optical Gravitational Lensing Experiment. OGLE-III Photometric Maps of the Large Magellanic Cloud},} ACTAA, 58, 89.
\newblock \doarXiv{0807.3889}

\bibitem[{I. {Vargas-Salazar} {et~al.}(2020){Vargas-Salazar}, {Oey}, {Barnes}, {Chen}, {Castro}, {Kratter}, \& {Faerber}}]{Vargas-Salazar2020}
{Vargas-Salazar}, I., {Oey}, M.~S., {Barnes}, J.~R., {et~al.} 2020, \bibinfo{title}{{A Search for In Situ Field OB Star Formation in the Small Magellanic Cloud},} \apj, 903, 42, \dodoi{10.3847/1538-4357/abbb95}

\bibitem[{I. {Vargas-Salazar} {et~al.}(2025){Vargas-Salazar}, {Oey}, {Eldridge}, {Weisserman}, {Januszewski}, {Becker}, {Zazzera}, {Castro}, {Kim}, {Kratter}, {Mateo}, \& {Bailey}}]{Vargas-Salazar2025}
{Vargas-Salazar}, I., {Oey}, M.~S., {Eldridge}, J.~J., {et~al.} 2025, \bibinfo{title}{{New Field OB and OBe Binaries of the SMC Wing: Observational Properties and Population Modeling},} arXiv e-prints, arXiv:2506.13004, \dodoi{10.48550/arXiv.2506.13004}

\bibitem[{J.~I. {Villase{\~n}or} {et~al.}(2021){Villase{\~n}or}, {Taylor}, {Evans}, {Ram{\'\i}rez-Agudelo}, {Sana}, {Almeida}, {de Mink}, {Dufton}, \& {Langer}}]{Villasenor2021}
{Villase{\~n}or}, J.~I., {Taylor}, W.~D., {Evans}, C.~J., {et~al.} 2021, \bibinfo{title}{{The B-type binaries characterization programme I. Orbital solutions for the 30 Doradus population},} \mnras, 507, 5348, \dodoi{10.1093/mnras/stab2197}

\bibitem[{D. {Zaritsky} {et~al.}(2004){Zaritsky}, {Harris}, {Thompson}, \& {Grebel}}]{Zaritsky2004}
{Zaritsky}, D., {Harris}, J., {Thompson}, I.~B., \& {Grebel}, E.~K. 2004, \bibinfo{title}{{The Magellanic Clouds Photometric Survey: The Large Magellanic Cloud Stellar Catalog and Extinction Map},} \aj, 128, 1606, \dodoi{10.1086/423910}

\end{thebibliography}

\end{document}